\documentclass[nofootinbib,aps,prl,reprint,groupedaddress,superscriptaddress,showpacs]{revtex4-1}
\usepackage[english]{babel}
\usepackage[utf8x]{inputenc}
\usepackage{amssymb}
\usepackage{amsthm}
\usepackage{amsmath}
\usepackage{hyperref}
\usepackage{subfigure}
\usepackage[percent]{overpic} 
\usepackage{picture}
\usepackage{color}
\usepackage{nth}
\usepackage{gensymb}
\usepackage{cleveref}
\usepackage{textcomp}
\usepackage{makecell}
\usepackage{lineno} 
\usepackage{textcomp}
\graphicspath{{img/}}

\begin{document}

\title{First emittance measurement of the beam-driven plasma wakefield accelerated electron beam}

\author{V. Shpakov}
\email[]{vladimir.shpakov@lnf.infn.it}
\author{M.P. Anania}
\author{M. Behtouei}
\author{M. Bellaveglia}
\author{A. Biagioni}
\affiliation{Laboratori Nazionali di Frascati, Via Enrico Fermi 54, 00044 Frascati, Italy}
\author{M. Cesarini}
\affiliation{Laboratori Nazionali di Frascati, Via Enrico Fermi 54, 00044 Frascati, Italy}
\affiliation{Sapienza University, Piazzale Aldo Moro 5, 00185 Rome, Italy}
\author{E. Chiadroni}
\affiliation{Laboratori Nazionali di Frascati, Via Enrico Fermi 54, 00044 Frascati, Italy}
\author{A. Cianchi}
\affiliation{University of Rome Tor Vergata and INFN, Via della Ricerca Scientifica 1, 00133 Rome, Italy}
\author{G. Costa}
\author{M. Croia}
\author{A. Del Dotto}
\author{M. Diomede}
\author{F. Dipace}
\author{M. Ferrario}
\author{M. Galletti}
\author{A. Giribono}
\author{A. Liedl}
\author{V. Lollo}
\author{L. Magnisi}
\affiliation{Laboratori Nazionali di Frascati, Via Enrico Fermi 54, 00044 Frascati, Italy}
\author{A. Mostacci}
\affiliation{Sapienza University, Piazzale Aldo Moro 5, 00185 Rome, Italy}
\author{G. Di Pirro}
\author{L. Piersanti}
\author{R. Pompili}
\author{S. Romeo}
\affiliation{Laboratori Nazionali di Frascati, Via Enrico Fermi 54, 00044 Frascati, Italy}
\author{A.R. Rossi}
\affiliation{INFN Milano, via Celoria 16, 20133 Milan, Italy}
\author{J. Scifo}
\author{C. Vaccarezza}
\author{F. Villa}
\affiliation{Laboratori Nazionali di Frascati, Via Enrico Fermi 54, 00044 Frascati, Italy}
\author{A. Zigler}
\affiliation{Laboratori Nazionali di Frascati, Via Enrico Fermi 54, 00044 Frascati, Italy}
\affiliation{Racah Institute of Physics, Hebrew University, 91904 Jerusalem, Israel}


\date{\today}

\begin{abstract}
Next-generation plasma-based accelerators can push electron beams to GeV energies within centimetre distances. The plasma, excited by a driver pulse, is indeed able to sustain huge electric fields that can efficiently accelerate a trailing witness bunch, which was experimentally demonstrated on multiple occasions. Thus, the main focus of the current research is being shifted towards achieving a high quality of the beam after the plasma acceleration. In this letter we present beam-driven plasma wakefield acceleration experiment, where initially preformed high-quality witness beam was accelerated inside the plasma and characterized. In this experiment the witness beam quality after the acceleration was maintained on high level, with $0.2\%$ final energy spread and $3.8~\mu m$ resulting normalized transverse emittance after the acceleration. In this article, for the first time to our knowledge, the emittance of the PWFA beam was directly measured.

\end{abstract}


\maketitle

The idea of plasma wakefield acceleration, proposed by Tajima and Dawson in 1979 \cite{tajima_dawson}, has attracted a lot of attention in recent years. So far several experiments have demonstrated plasma-based acceleration using both laser pulses~\cite{leemans2006gev,geddes2004high,faure2004laser,deng2019generation} or charged particle beams~\cite{2007Natur.445..741B,litos2014high,adli2018acceleration,loisch2018observation} as drivers, demonstrating the possibility to generate large accelerations of about tens of GV/m, i.e. orders of magnitude larger than what can be provided by conventional radio-frequency (RF) accelerators. Such increase of the acceleration gradient allows to build ultra-compact, down to table-top, accelerators that can be highly beneficial for many applications including advanced radiation sources based on  Free Electron Lasers (FEL)~\cite{ackermann2007operation,emma2010first,petrillo2013observation}, Compton scattering~\cite{schoenlein1996femtosecond,bacci2013electron,jochmann2013high}, THz radiation~\cite{chiadroni2020versatile,giorgianni2016tailoring}, and a wide range of medical and industrial applications.

In last several years laser driven wakefield accelerated (LWFA) electron beams were successfully used to demonstrate production of radiation from undulators~\cite{schlenvoigt2008compact,andre2018control,fuchs2009laser,ghaith2019tunable,maier2020water} or Thomson scattering~\cite{khrennikov2015tunable}. However, the further advancement of such plasma based sources, like FEL radiation, requires substantial improvement of the beam quality in terms of energy spread and emittance~\cite{pompili2020energy,wang2016high}. In this letter we report the results of a beam-driven plasma wakefield acceleration (PWFA)~\cite{sprangle1996laser,hogan2010plasma,hidding2019fundamentals} experiment conducted at the SPARC\_LAB test-facility \cite{ferrario2013sparc_lab}. Initially preformed by a RF-linac a high quality witness beam was accelerated in plasma and characterized. Using a technique for energy spread minimization described in~\cite{pompili2020energy} we have achieved resulting energy spread of the PWFA beam of the order of $0.2\%$. Such energy spread allowed us to transport the beam through a conventional transfer line and use standard multi-shot diagnostics based on quadrupole-scan. The normalized emittance~\cite{cianchi2015six} of the PWFA beam was measured at the level of $3.8$~\textmu m. The experimental results are supported by complete start-to-end simulations. These results demonstrate the possibility to use PWFA beams in the frame of a conventional accelerator to pilot user applications.

The SPARC\_LAB photo-injector consists of a 1.6-cell S-band electron gun, followed by two $3$~m long S-band accelerating sections and one $1.4$~m C-band section.
Both driver and the witness beams are generated directly from the photo-cathode using the laser-comb technique~\cite{villa2014laser,2011NIMPA.637S..43F}  by illuminating it with two consecutive ultra-short laser pulses (130~fs, rms), whose delay can be adjusted through an optical delay-line.
A first S-band section is used as a bunch compressor by means of the velocity-bunching technique~\cite{serafini2001velocity,ferrario2010experimental}, allowing in turn to both accelerate and compress the beam. Moreover, it allows for a  precise adjustment of the bunch duration and distances~\cite{pompili2016beam} and, with respect to other methods employing masks or scrapers in dispersing sections~\cite{litos2014high,roussel2020single}, there is no loss of charge.

\begin{figure*}[t]
\centering
\includegraphics[width=0.98\linewidth]{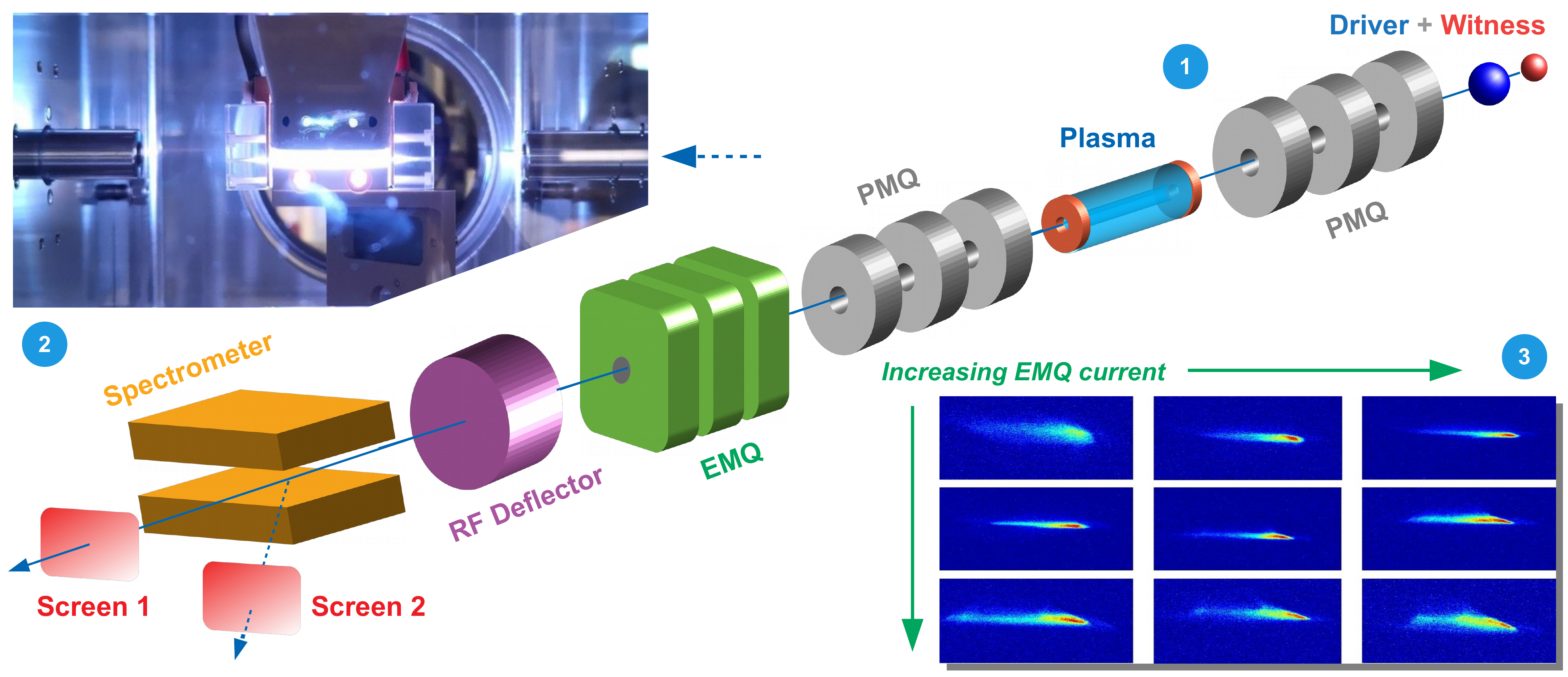}
\caption{Experimental setup. (1) The incoming driver and witness bunches are focused by a triplet of permanent-magnet quadrupoles (PMQ) into the plasma accelerator module. A second triplet of PMQs is used to extract and transport the bunches up to two diagnostics stations located on the straight (straight screen, Screen 1) and bent (spectrometer screen, Screen 2) beam lines. A triplet of electromagnetic quadrupoles (EMQ), a RF-Deflector device and a magnetic spectrometer allow to completely characterize the beam. (2) The plasma module consists of a 3~cm-long capillary where the plasma is produced by ionizing H$_2$ gas with an high-voltage discharge. (3) The inset shows a typical quadrupole-scan of the plasma accelerated witness performed on the spectrometer screen by varying the EMQ currents to reconstruct its normalized emittance.}
\label{CapillarySetup}
\end{figure*}

The plasma experimental setup, depicted in Fig.~\ref{CapillarySetup}, is installed downstream from the photo-injector. The plasma is contained in a 3~cm long 3D-printed plastic capillary and it is produced by ionizing Hydrogen gas, injected through two inlets, by means of a high-voltage discharge (12~kV, 300~A) at 1~Hz repetition rate. The plasma density is measured using Stark-broadening based diagnostics~\cite{filippi2016spectroscopic}. The mean value of the plasma density is controlled by delaying the beam arrival time with respect to the discharge as reported in \cite{shpakov2019longitudinal}. The beam is focused at the entrance of the capillary by a triplet of permanent magnet quadrupoles (PMQs)~\cite{pompili2018compact}. After the interaction with the plasma, the beam is transported up to the spectrometer using a second triplet of PMQs and three electromagnetic quadrupoles (EMQs). The diagnostics is completed by an RF-Deflector to characterize the longitudinal beam profile. The energy spectrum is finally measured with a Ce:YAG scintillator screen (spectrometer screen, Fig.\ref{CapillarySetup}) located downstream the magnetic spectrometer.

The beam configuration consists of a 200~pC driver with $86.2\pm0.1$~MeV energy (0.2~MeV energy spread) and $250$~fs rms duration followed, at a distance $1.0$~ps (with the jitter of this distance $\sim30$~fs), by a 20~pC witness with $E_i=85.6\pm0.2$~MeV energy (0.24~MeV energy spread) and $\sigma_t=35$~fs rms duration, corresponding to about 570~A peak current. The driver and witness bunches are then focused down to $\approx 25$~\textmu m and $\approx 13\mu m$ correspondingly, and injected into the plasma with density $n_p\approx 1.5\times10^{15}$~cm$^{-3}$, obtained by delaying the beam time of arrival with respect to the discharge trigger. The fluctuation of the plasma density was measured at $11\%$. The witness energy spread and emittance prior to the plasma module are $\sigma_E=0.24\pm0.10$~MeV and $\epsilon_{n,y}=2.82\pm0.56$~\textmu m, respectively. During the experiments the charge of the witness was controlled using the beam current monitor before plasma and the scintillator screen after the plasma. The CCD camera counts of the witness beam with and without plasma were compared and no loss of the charge was detected. 

Considering these parameters, the experiment is thus carried out in the \textit{quasi-nonlinear} (QNL) regime~\cite{rosenzweig2010plasma}, where the driver bunch density exceeds the plasma one and induces blowout process but, due to its relatively small charge, the produced perturbation is linear.
Here we are defining $\widetilde{Q}=N_b k_p^3/n_p$ as the normalized bunch charge that quantifies the plasma response, with $N_b$ the number of electrons contained in the driver and $k_p$ the plasma wave-number. The QNL regime, described for the proposed configuration ($\widetilde{Q}\approx 0.37$), is characterized by $\widetilde{Q}<1$, contrary to the linear ($\widetilde{Q}\ll 1$) and nonlinear (or blowout, $\widetilde{Q}>1$) cases.

\begin{figure}[ht]
\centering
\includegraphics[width=0.98\linewidth]{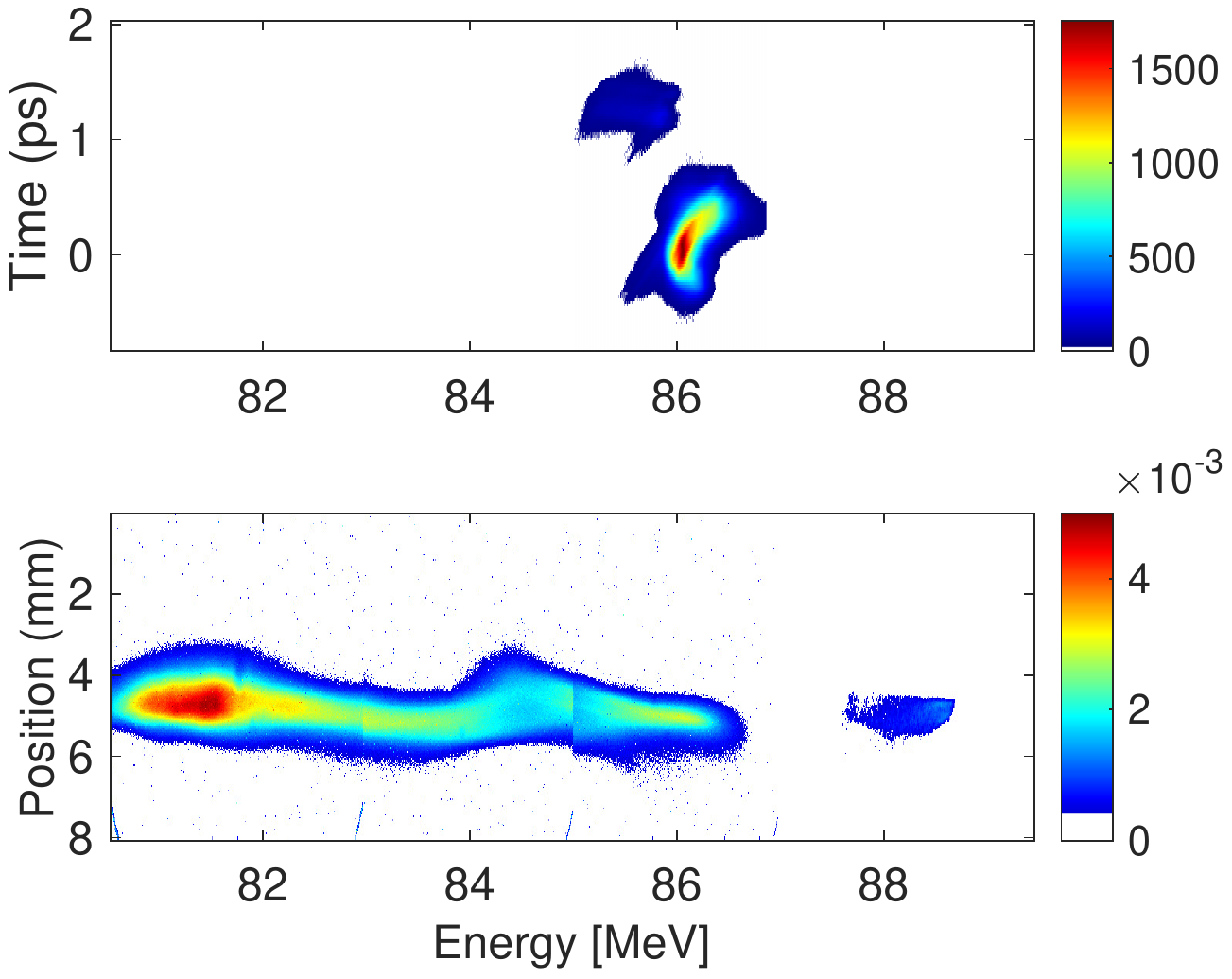}
\caption{Longitudinal phase-space of the driver and witness bunches upstream (top) and bunch train energy spectrum downstream the plasma accelerator module (bottom). To reconstruct the energy-depleted spectrum of the driver after witness acceleration, the  bottom plot is obtained by merging several images acquired for different currents of the magnetic spectrometer.}
\label{LPSimage}
\end{figure}

\begin{figure}[ht]
\centering
\includegraphics[width=0.98\linewidth]{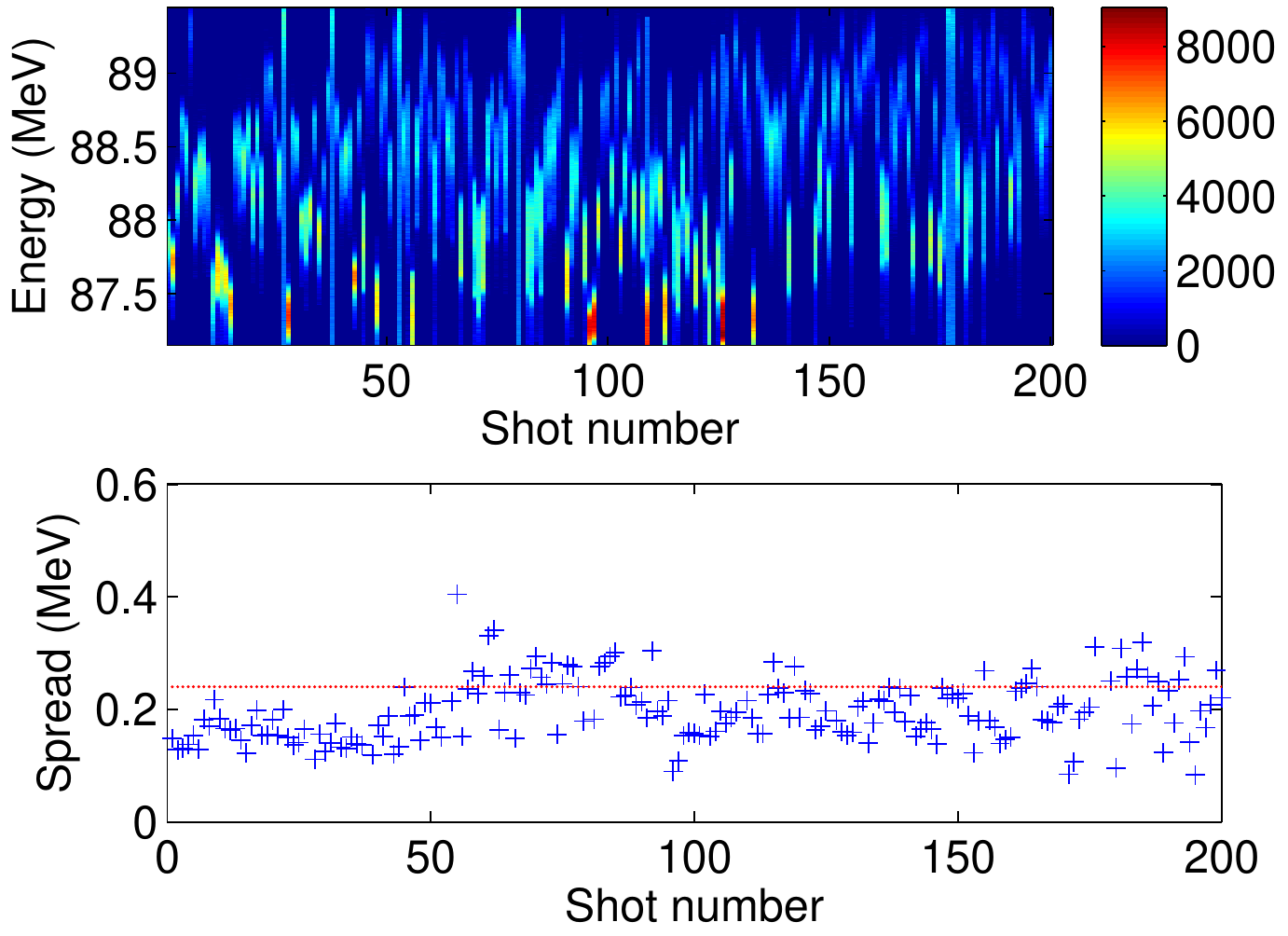}
\caption{(top) Energy spectrum traces of 200 shots of the witness bunch after acceleration in the plasma. (bottom) Analysis of these shots in terms of energy spread. The energy spread of the witness beam before the acceleration is also reported (red dotted line).}
\label{EnergyJitter}
\end{figure}

The longitudinal phase-space (LPS) of the two bunches before acceleration is shown in Fig.~\ref{LPSimage} (top image). The energy spectrum of the bunch train after the acceleration is depicted in the bottom image of the Fig.~\ref{LPSimage}. The energy window of the spectrometer is about $2$ MeV, therefore to reconstruct the energy-depleted spectrum of the driver (spanning $\approx 6$~MeV) several images have been acquired for different currents of the magnetic spectrometer, and then merged together (bottom plot of Fig.\ref{LPSimage}). The plot highlights a 3~MeV acceleration of the witness in the plasma, reaching a final energy of $88.6\pm0.5$~MeV, corresponding to $\approx 100$~MV/m acceleration gradient. Given that the energy fluctuation before the acceleration was $\sim0.2$~MeV, the increase of the energy jitter (up to $\sim0.5$~MeV) was fully attributed to the plasma. Here we assume that the jitter from the linac and the plasma are independent and total jitter is $\Delta E^{2}_{total}=\Delta E^{2}_{linac}+\Delta E^{2}_{plasma}$. Thus, with $\sim3$~MeV as typical energy boost, we have evaluated instability of the plasma acceleration at $\sim15\%$. Figure~\ref{EnergyJitter} shows the witness energy spectra of 200 shots. The resulting energy and energy spread was computed for each of the shots in the series. This series was taken continuously, but the $~15\%$ of shots with the beam out of the spectrometer screen has been excluded. Such empty shots are attributed to misfire of the gas/discharge system. The small energy spread of the beam is a paramount characteristics for high quality beams, thus in this work was employed a recently developed technique~\cite{pompili2020energy}. By using a combination of the positive energy-chirp (larger energy particles on the head of the bunch) and beam-loading effects we were able to mitigate any energy spread growth, but also to achieve a slight reduction of the total energy spread by removing, partially, the correlated one. The final energy spread after the acceleration was $\sigma_E=0.21\pm0.12$~MeV. It needs be highlighted that the method used here to keep under control the energy spread, can be scaled to a larger energy gain by using larger plasma density, although correspondingly larger energy chirp will be also required. For more information see Extension 1.

\begin{figure}[ht]
\centering
\subfigure{
\begin{overpic}[width=0.98\linewidth]{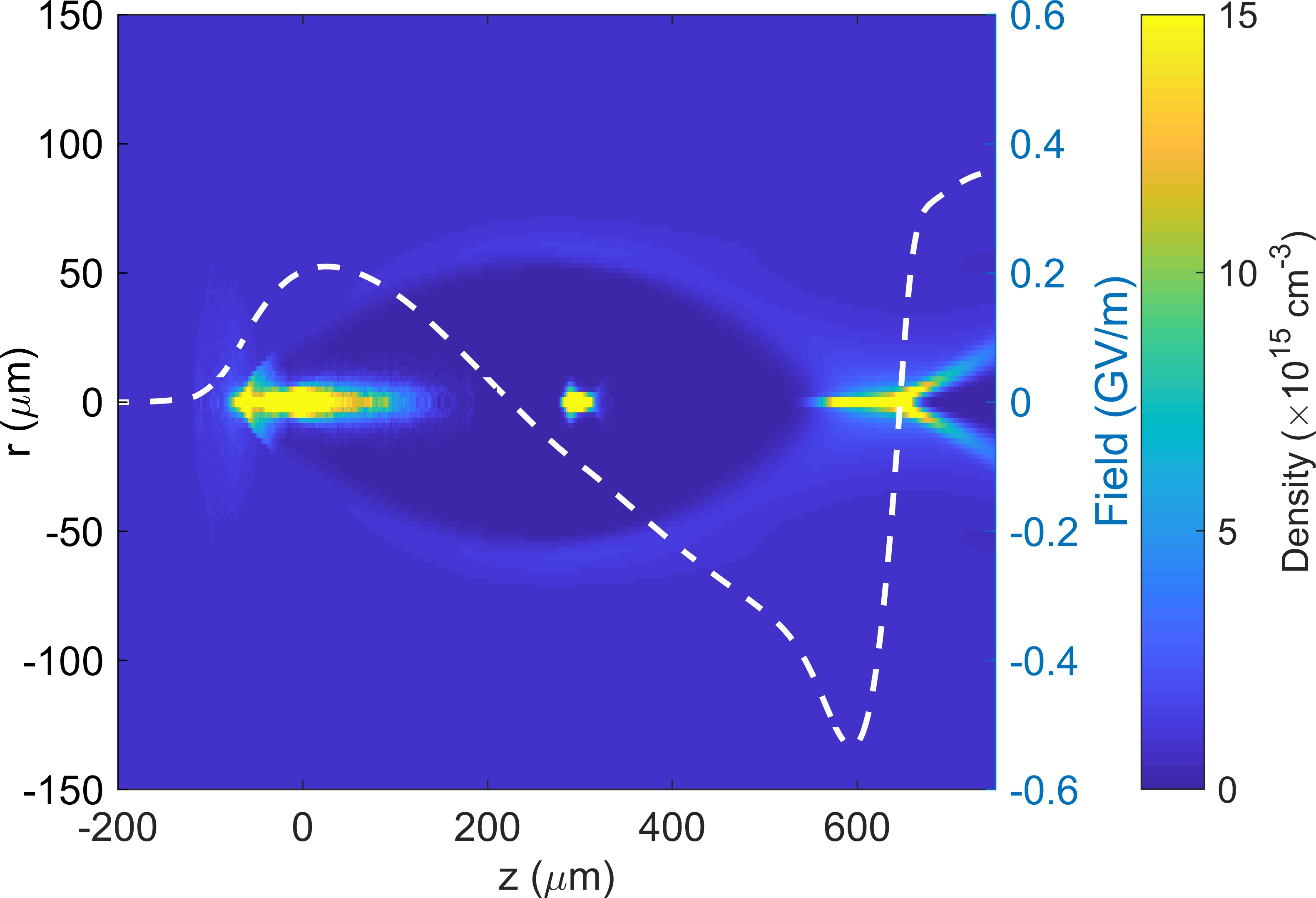}
\put(12,62){\color{white}\textbf{a}}
\end{overpic}
\label{ArchitectDensity}
}
\subfigure{
\begin{overpic}[width=0.97\linewidth]{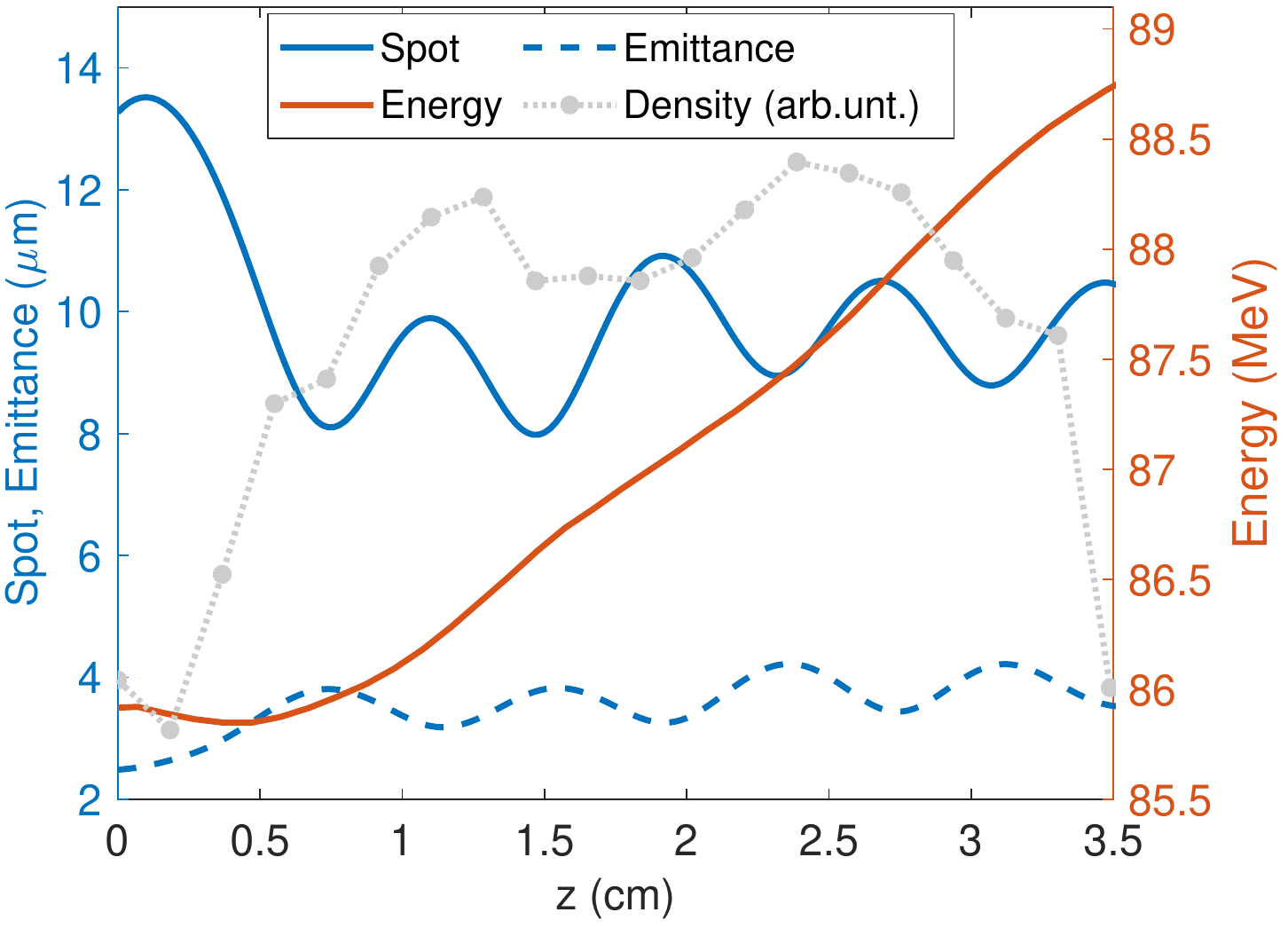}
\put(12,57){\color{black}\textbf{b}}
\end{overpic}
\label{EmittanceArchitect}
}
\caption{The evolution of the beam parameters along the plasma section. On the image a) is depicted 2-D distribution of the plasma during the acceleration (the beam is traveling from right to left) and the strength of the longitudinal electric field (white dash line). The evolution of the beam parameters along the plasma is depicted on the image b).}
\label{mainScan}
\end{figure}

The transformer ratio is defined as the peak accelerating field divided by the peak decelerating field \cite{fmassimo}. In our case it was estimated from simulations (see Fig.\ref{ArchitectDensity}) and it is of the order of 2.5. The witness is located quite far from the rear of the blowout region (where the acceleration is maximized) to avoid an over-rotation of its LPS and, in turn, a large growth of its energy spread. This would make its detection quite challenging also due to the limited energy acceptance of our spectrometer ($\sim2~MeV$). Therefore, in order to achieve a small witness energy spread, the witness is placed not at the far rear of the blowout region but close to its center. Our aim, indeed, is not to demonstrate the largest possible acceleration but to preserve as much as possible the witness quality (spread, emittance) in order to make it measurable (e.g. by quadscan). 

\begin{figure}[ht]
\centering
\includegraphics[width=0.98\linewidth]{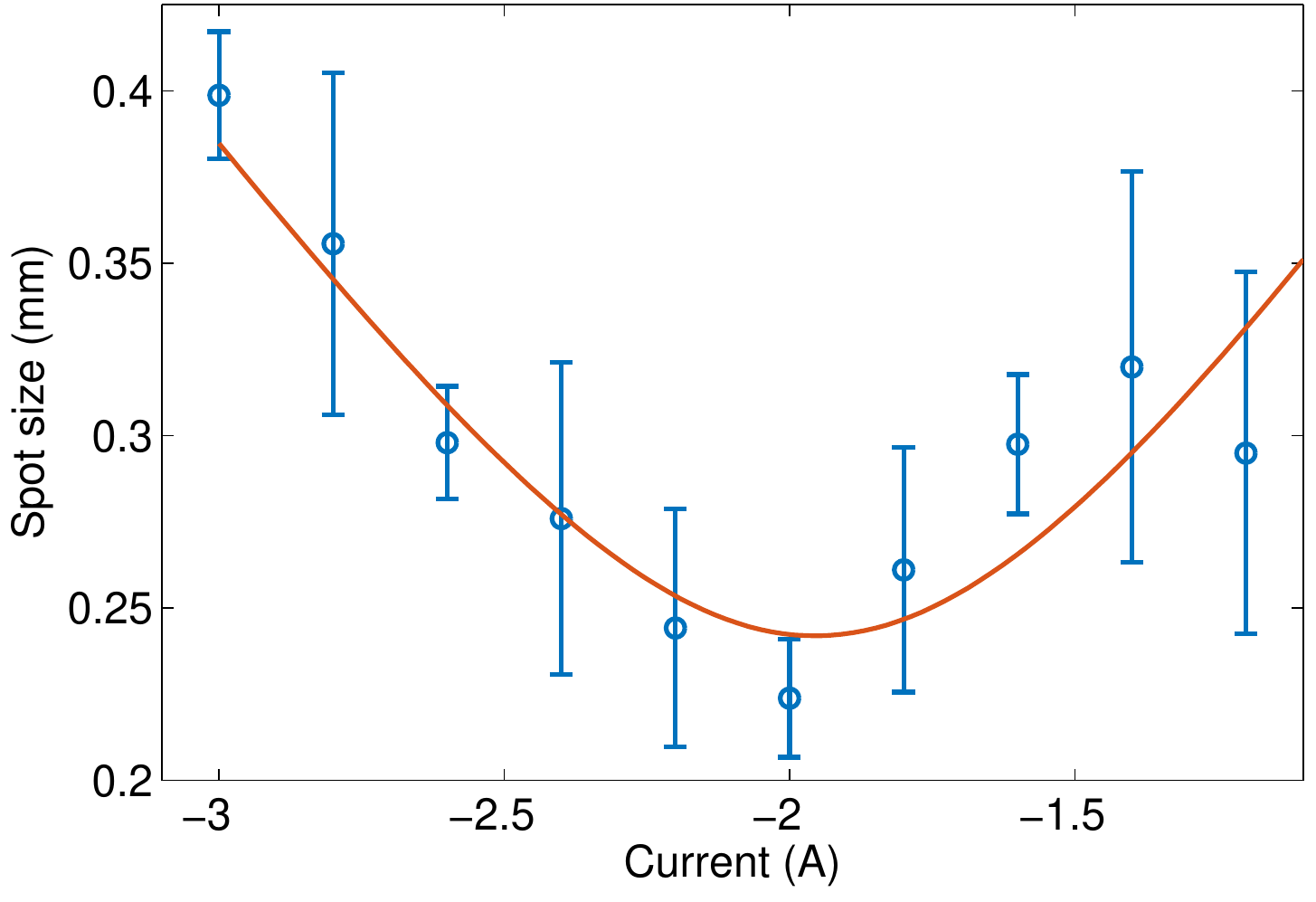}
\caption{Fit of the quadrupole scan for the emittance measurement of the PWFA beam. Each point is the rms value of the vertical beam size, taken over 10 shots.}
\label{EmittanceFit}
\end{figure}

The obtained low energy spread is of paramount importance when considering its effects on the resulting normalized emittance.
Downstream the PWFA module, indeed, there could be a rapid degradation of the witness emittance over the drift $s\approx 20$~cm to reach the second PMQ triplet according to the relation~\cite{migliorati2013intrinsic}
\begin{equation}
    \label{normEmittanceEQ}
    \epsilon_{n,f}^2= \epsilon_{n,i}^2 + \gamma^{2}{{\sigma_E^2}\over E^2}\sigma_{x^\prime}^4 s^2~,
\end{equation}
where $\epsilon_{n,i}$ and $\epsilon_{n,f}$ are the initial and final normalized emittances, $\gamma$ the relativistic Lorentz factor and $\sigma_{x^\prime}\approx 2.4$~mrad the beam divergence (was estimated from simulations). We can see that the contribution of the energy spread to the final normalized emittance (second term of Eq.~\ref{normEmittanceEQ}) is almost negligible, so that its value does not change during the transport downstream the PWFA module.
Several methods can be employed to retrieve the normalized emittance. Previous experiments, for instance, estimated the emittance in a single-shot way by sampling the vertical size of the beam as a function of the energy~\cite{barber2018parametric,weingartner2012ultralow}. However, these techniques would require rather large energy spreads ($\gg 1\%$) and it is thus not applicable to our case and, in general, to very low energy spread beams. For such a reason and considering also the high stability of the accelerated witness, we used the classical quadrupole scan technique to estimate its emittance \cite{mostacci2012chromatic}. The measurement is performed on the screen downstream from the spectrometer, where, due to the difference in energy, the two beams are well separated. The scan consists in measuring the witness vertical spot size as a function of the current used in the EM quadrupoles upstream. The resulting measurement is depicted in Fig.\ref{EmittanceFit}. By performing a numerical fit on the experimental value, the resulting normalized emittance is $\epsilon_n=3.8\pm0.5~\mu m$. Being such a value not affected by the transport downstream from the plasma, we conclude that the emittance increased only during the acceleration process. At the current setup the horizontal emittance can not be measured, thus the brightness of the beam was evaluated assuming that the beam has similar emittance in both planes. Under such conditions the resulting beam brightness can be estimated at the level of $\sim10^{13}~A/m^2$\cite{cianchi2017observations}.

To support the experimental observations, we performed a complete start-to-end simulation using Architect~\cite{marocchino2016efficient,alberto_marocchino_2016_49572,massimo2016comparisons}, a hybrid code where the electron bunches are treated with a kinetic approach as in a Particle-in-Cell code while the background plasma is simulated as a fluid.
Figure~\ref{ArchitectDensity} shows a snapshot obtained at half the capillary ($z=1.5$~cm) of the two bunches travelling through the plasma. The longitudinal electric field resulting is also reported (white dashed line). The main quantities related to the witness bunch are highlighted in Fig.~\ref{EmittanceArchitect}. For the simulation the experimentally measured parameters of the beam (e.g. LPS, emittance), which were indicated above, were directly imported to be used in calculations. The used plasma density profile is shown in Fig.\ref{EmittanceArchitect}, was also measured experimentally \cite{biagioni2016electron} and imported to the code. The density profile has an average value in the center (inside of the capillary) $n_p\approx 1.5\times10^{15}$~cm$^{-3}$, as was indicated above. Simulations have been performed with a longitudinal resolution of 2 $\mu m$ and a transverse resolution of 1 $\mu m$; reasonable being the plasma wavelength about 834 $\mu m$ that values also keep the computational time small. The advancing time step is 1.2 fs, and a correct sampling of both driver and witness is also guaranteed. The simulated acceleration, as well as the beam emittance are in agreement (within the error margin) with the experimental data. The position of the witness beam, that was used in the experiment ($\sim1.0~ps$ from the driver), represents a best compromise between the acceleration and energy spread compensation. We also can see a clear oscillations of the witness envelope along the propagation distance as well as of its emittance. The growth of the emittance by approximately $37\%$ is mainly due to an mismatched witness spot size~\cite{litos2019beam}.

In conclusion, in this work we presented the results of a beam driven PWFA experiments where the high-quality preformed witness beam was injected inside of the plasma and accelerated. Thanks to the achieved low energy spread after the acceleration $\approx 0.23\%$, with $\approx 0.28\%$ before the acceleration, we were able, using a conventional transport line and multi-shot quadrupole scan technique, to measure the transverse emittance of the beam. The final normalized emittance of the beam was measured at the level of $3.8$~\textmu m, with initial emittance $2.8$~\textmu m. The reported simulation studies indicate that such a growth was mainly caused by non-optimal transverse matching conditions. Thus in this experiment we were able to accelerate the beam inside of the plasma and, in general, preserve its quality afterwards.         

One of the ultimate goals of the PWFA experiments is to provide an electron beam suited for a wide range of applications. Capability to use plasma-accelerated beams with conventional/existing infrastructure can be an important milestone towards achieving such a goal. At SPARC\_LAB the plasma module is an insertion into a pre-existing machine and in this work we have demonstrated that such module can be operated as an integral part of a conventional accelerator, like using a quadrupole scan to characterize the beam. These results represent a fundamental step towards the realization of future compact accelerators providing high-quality electron beams for user-oriented applications like Free-Electron Lasers.

\section*{Acknowledgments}

\begin{acknowledgments}
This work has been partially supported by the EU Commission in the Seventh Framework Program, Grant Agreement 312453-EuCARD-2, the European Union Horizon 2020 research and innovation program, Grant Agreement No. 653782 (EuPRAXIA) and the INFN with the GRANT73/PLADIP grant. This work has been partially funded by the 5th National Scientific Committee of the INFN with the SL\_COMB2FEL experiment. The authors thank D. Pellegrini for the realization of the HV discharge pulser, M. Del Franco for providing the layout of the SPARC\_LAB photo-injector, and INFN Accelerator Division operators for the help with machine operation.

\end{acknowledgments}

\bibliography{biblio}
\bibliographystyle{apsrev4-1}

\end{document}